# RADIATIVE NEUTRON CAPTURE ON $^{14}$C AND $^{14}$N

**N. V. Afanasyeva[1,2,*], S. B. Dubovichenko[1,3,†] and
A. V. Dzhazairov-Kakhramanov[1,3,‡]**

1   V. G. Fessenkov Astrophysical Institute "NCSRT" NSA RK,
    Observatory 23, Kamenskoe plato, 050020, Almaty, Kazakhstan.
2   Al-Farabi Kazakh National University,
    av. Al-Farabi 71, 050040, Almaty, Kazakhstan.
3   Institute of Nuclear Physics of the NNC RK,
    str. Ibragimova 1, 050032, Almaty, Kazakhstan.

**Abstract**
The possibility of description of the experimental data on total cross sections of the radiative neutron capture on $^{14}$C and $^{14}$N is considered within the frame of the potential cluster model with forbidden states and their classification according to Young schemes. It is shown that the using model and the potential construction methods allow to reproduce correctly the behavior of experimental cross sections at the energies from 10 meV ($10^{-2}$ eV) to 1 MeV.

**PACS (2008):** 21.60.Gx, 25.20.-x, 25.40.Lw, 26.35.+c, 26.

## 1. Introduction

Although, the radiative neutron capture reactions on $^{14}$C and $^{14}$N at astrophysical energies do not take part in the main thermonuclear cycles [1–3] directly, they can play a certain role in some models of the Big Bang [4], where it is considered that the primordial nucleosynthesis has the main reaction chain of the form

$^{1}$H(n,γ)$^{2}$H(n,γ)$^{3}$H($^{2}$H,n)$^{4}$He($^{3}$H,γ)$^{7}$Li(n,γ)$^{8}$Li($^{4}$He,n)$^{11}$B(n,γ)$^{12}$B(β$^{-}$)$^{12}$C(n,γ)$^{13}$C(n,γ)
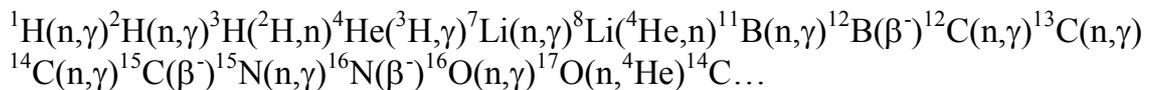
$^{14}$C(n,γ)$^{15}$C(β$^{-}$)$^{15}$N(n,γ)$^{16}$N(β$^{-}$)$^{16}$O(n,γ)$^{17}$O(n,$^{4}$He)$^{14}$C…

etc. [5], where the first of these capture processes is part of, and not only at low energies, that are considered later. Here, we will dwell on the n$^{14}$C → $^{15}$Cγ radiative capture reaction at energies 23 keV–1.0 MeV, and then the n$^{14}$N → $^{15}$Nγ capture reaction, which leads to the formation and accumulation of the $^{15}$N nuclei, will be considered. Thereby, this reaction is additional to the $^{14}$C(n,γ)$^{15}$C(β$^{-}$)$^{15}$N process, which take place in the main reaction chain and increases the amount of $^{15}$N, taken part in subsequent synthesis reactions of heavier elements.

    The potentials of the scattering processes, within the frame of using potential

---

* E-mail: n.v.afanasyeva@gmail.com
† E-mail: dubovichenko@gmail.com
‡ E-mail: albert-j@yandex.ru

cluster model (PCM) [6,7], usually are constructed on the basis of description of the elastic scattering phase shifts taking into account their resonance behavior, derived at the phase shift analysis from the experimental differential cross sections. The interactions of the bound states (BS), except the description of the scattering phase shifts, are clarified from the additional reconstruction requirement of some ground state (GS) characteristics of nucleus – mean square radius, asymptotic constant (AC), binding energy in the considered channel. Thereby, it is supposed that such BS is caused, in general, by the cluster channel consisting of initial particles, which take part in the reaction [8].

## 2. Calculation methods

The nuclear part of the intercluster interaction potential, for carrying out the calculations of photonuclear processes in the considered cluster systems, has the form:

$$V(r) = -V_0 \exp(-\alpha r^2), \qquad (1)$$

with zero Coulomb potential.

The asymptotic constant for any GS potential was calculated using the asymptotics of the wave function (WF) having a form of exact Whittaker function [9]

$$\chi_L(R) = \sqrt{2k}\, C_W W_{-\eta_L + 1/2}(2kR),$$

where $\chi_L$ is the numerical wave function of the bound state obtained from the solution of the radial Schrödinger equation and normalized to unity; $W$ is the Whittaker function of the bound state which determines the asymptotic behavior of the WF and represents the solution of the same equation without nuclear potential, i.e., long distance solution; $k$ is the wave number determined by the channel binding energy; $\eta$ is the Coulomb parameter that is equal to zero in this case; $L$ is the orbital moment of the bound state.

The variational method (VM) with independent variation of all parameters and with expansion of cluster wave function on the non-orthogonal Gaussian basis was used for additional control of the nuclear binding energy calculations

$$\Phi_L(R) = \frac{\chi_L(R)}{R} = R^L \sum_i C_i \exp(-\gamma_i R^2),$$

where $\gamma_i$ – variational parameters and $C_i$ – WF expansion coefficients.

The total radiative capture cross sections $\sigma(NJ, J_f)$ for $EJ$ and $MJ$ transitions in the case of potential cluster model are given, for example, in works [10–12] and have the following form:

$$\sigma_c(NJ, J_f) = \frac{8\pi K e^2}{\hbar^2 q^3} \frac{\mu}{(2S_1+1)(2S_2+1)} \frac{J+1}{J[(2J+1)!!]^2} \times$$
$$\times A_J^2(NJ, K) \sum_{L_i, J_i} P_J^2(NJ, J_f, J_i) I_J^2(J_f, J_i)$$

where σ – total radiative capture cross section; μ – reduced mass of colliding particles; q – wave number in initial channel; $S_1$, $S_2$ – spins of particles in initial channel; K, J – wave number and momentum of γ-quantum; N – it is E or M transitions of J multipole order from the initial $J_i$ to the final $J_f$ nucleus state.

The value $P_J$ for electric orbital EJ(L) transitions ($S_i = S_f = S$) has the form [12]

$$P_J^2(EJ, J_f, J_i) = \delta_{S_i S_f} \left[(2J+1)(2L_i+1)(2J_i+1)(2J_f+1)\right](L_i 0 J 0 | L_f 0)^2 \begin{Bmatrix} L_i & S & J_i \\ J_f & J & L_f \end{Bmatrix}^2,$$

$$A_J(EJ, K) = K^J \mu^J \left(\frac{Z_1}{m_1^J} + (-1)^J \frac{Z_2}{m_2^J}\right),$$

$$I_J(J_f, J_i) = \langle \chi_f | R^J | \chi_i \rangle.$$

Here $L_f$, $L_i$, $J_f$, $J_i$ – angular and total moments in initial (*i*) and final (*f*) channels; $m_1$, $m_2$, $Z_1$, $Z_2$ – masses and charges of the particles in initial channel; $I_J$ – integral over wave functions of initial $\chi_i$ and final $\chi_f$ states, as functions of cluster relative motion with intercluster distance R.

Let us stress that we never used such notation as a spectroscopic factor (see, for example, [11]), i.e., its value takes equal to unit. Apparently, now it is not needed to enter the additional factor $S_f$ [12] under consideration the capture reaction in the PCM for potentials agreed in continuous spectrum with the characteristics of scattering processes that take into account the resonance phase shift shapes and, in discrete spectrum, where they describe the basic characteristics of the GS or BS of nucleus. Consequently, all attendant reaction effects, expressing in certain factors and coefficients, are taken into account by interaction potentials, because they are constructed on the basis of the description of observed, i.e., experimental characteristics of interacting clusters in the initial channel during capture and formed in the final state certain nucleus, with its description by the cluster structure consisting of the initial particles.

Because, at low energies the calculated cross section is practically straight line, it may be approximated by simple function of the form:

$$\sigma_{ap}(\mu b) = \frac{A}{\sqrt{E_n (keV)}}. \tag{2}$$

The value of the presented constant A, expressed in μb·keV$^{1/2}$, usually was determined by a single point of cross sections with minimal energy. It is possible to consider the absolute value of the relative deviation of the calculated theoretical cross sections and the approximation of this cross section by the function of Eq. (2) at the energy range from $10^{-5}$ to 100 eV:

$$M(E) = |[\sigma_{ap}(E) - \sigma_{theor}(E)] / \sigma_{theor}(E)|, \tag{3}$$

which usually is not exceed 1.0–2.0% at low energies.

The exact mass values ($m_n$ = 1.00866491597 amu) [13, 14] were specified in our calculations, and constant $\hbar^2/m_0$ takes equal to 41.4686 MeV fm$^2$.

## 3. Total cross sections of the neutron capture on $^{14}$C

Starting the analysis of total cross sections of the neutron capture on $^{14}$C and $^{14}$N with the formation of $^{15}$C and $^{15}$N in GS, let us note that the classification of orbital states of $^{14}$C in the n$^{13}$C system or $^{14}$N in the p$^{13}$C channel according to Young schemes was considered by us earlier in works [15, 16]. However, we regard the results on the classification of $^{15}$C and $^{15}$N nuclei by orbital symmetry in the n$^{14}$C and n$^{14}$N channels as the qualitative ones as there are no complete tables of Young schemes productions for the systems with a number of nucleons more than eight [17], which have been used in earlier similar calculations [6, 18, 19]. At the same time, just on the basis of such classification, we succeeded with description of available experimental data on the radiative $^{13}$C(n, γ)$^{14}$C [15] and $^{13}$C(p, γ)$^{14}$N capture processes [16]. Therefore, here we will use the classification procedure of cluster states by orbital symmetries, which leads us to certain number of FS and AS in partial intercluster potentials, so, to certain number of nodes for wave function of cluster relative motion.

Furthermore, let us suppose that it is possible to use the orbital Young scheme {4442} for $^{14}$C, therefore we have {1} × {4442} → {5442} + {4443} for the n$^{14}$C system within the frame of 1$p$ shell [17]. The first of the obtained schemes compatible with orbital moments $L = 0, 2$ and is forbidden, because there can not be five nucleons in the $s$ shell, and the second scheme is allowed and compatible with the orbital moment $L = 1$ [20]. Thus, limiting oneself only by lowest partial waves with orbital moments $L = 0$ and 1, it is possible to say that there are forbidden and allowed states in the $^2S_{1/2}$ potential for the n$^{14}$C system. The last of them corresponds to the GS of $^{15}$C with $J^\pi = 1/2^+$ and lays at the binding energy of the n$^{14}$C system –1.21809 MeV [21]. At the same time the potential of the $^2P$ elastic scattering waves has not FS. In the case of the n$^{14}$N system, the forbidden state is in the $^2S_{1/2}$ elastic scattering wave, and the $^2P_{1/2}$ wave has only AS, which is at the binding energy of the n$^{14}$N system –10.8333 MeV [21].

Now we will consider the radiative neutron capture process on $^{14}$C at energies from 20 keV to 1 MeV, approximately, where there are experimental data that are taken from the Moscow State University (MSU) data base [14]. The value obtained on the basis of resonance data at 3.103(4) MeV relative to the GS of $^{15}$C with $J^\pi = 1/2^-$, i.e., approximately higher by 1.9 MeV in the center-of-mass system (c.m.) than threshold of the n$^{14}$C channel and with the width about 40 keV [21] can be used for the $^2P_{1/2}$ wave potential of the n$^{14}$C scattering without FS. Because, we are studying only the energy range not far than 1.0 MeV, so it can be considered that the $^2P_{1/2}$ scattering phase shift simply equals zero. Consequently, the potential depth $V_0$ can be equaled to zero, because it has not forbidden states. The same applies to the $^2P_{3/2}$ scattering wave potential too, since the relevant resonance of $^{15}$C nucleus lays at higher energy value of 4.66 MeV.

The potential of the $^2S_{1/2}$ bound state with one FS has to correctly reproduce ground state binding energy of $^{15}$C with $J^\pi = 1/2^+$ in the n$^{14}$C channel at -1.21809 MeV [21] and reasonably describes the mean square radius of $^{15}$C, which value, apparently, should not considerably exceed the radius of $^{14}$C that is equal to 2.4962(19) fm [21]. Consequently, the next parameters were obtained:

$$V_{GS} = 93.581266 \text{ MeV}, \quad \alpha_{GS} = 0.2 \text{ fm}^{-2}. \tag{4}$$

The potential leads to the binding energy of –1.2180900 MeV at the accuracy of using finite-difference method (FDM) [22] equals $10^{-7}$ MeV, to the mean square charge radius $R_{ch}$ = 2.52 fm and mass radius of 2.73 fm. The zero value was used as neutron charge radius and its mass radius has taken equal to proton radius [13]. The value of 1.85(1) was obtained for the asymptotic constant at the range 7–27 fm. The error of the constant is determined by its averaging over the above mentioned distance.

The value of 1.13 $fm^{-1/2}$ for AC was given in review [23] with the reference to [24], that gives 1.65 after recalculation to the dimensionless quantity at $\sqrt{2k} = 0.686$. The detailed review of the values of this constant is given in one of the latest work [25] devoted to the determination of the AC from the characteristics of different reactions – its value is in the limit from 1.22(6) $fm^{-1/2}$ to 1.37(5) $fm^{-1/2}$, that gives 1.8–2.0 after recalculation. In work [25] itself, the values of AC were obtained from 1.25 $fm^{-1/2}$ to 1.52 $fm^{-1/2}$ or after recalculation in the interval 1.8–2.2 with the recommended value 1.87(6), which practically does not differ from the value that we have obtained.

The variational method (VM) [22] was used for the additional control of calculations of the GS energy of $^{15}$C, which with the dimension of the basis $N = 10$ and independent variation of the potential parameters for the BS potential of Eq. (4) allowed to obtain the energy of –1.2180898 MeV. The parameters of the radial WF are listed in Table 1 and residuals have the order of $10^{-12}$ [22]. The asymptotic constant at the range 10–25 fm and the charge radius do not differ from the previous results of the FDM. Since, the variational energy decreases as the dimension of the basis increases and yields the upper limit of the true binding energy [6], and the finite-difference energy increases as the step size decreases and number of steps increases [22], the average value of –1.2180899(1) MeV can be taken as a realistic estimate of the binding energy in this potential. Therefore, the real accuracy of determination of the GS binding energy of $^{15}$C in the n$^{14}$C cluster channel for the potential of Eq. (4), using two methods (FDM and VM) and two different computer programs, is at the level ±0.1 eV.

**Table 1:** The coefficients and expansion parameters of the radial variational wave function of the n$^{14}$C system for the GS potential of Eq. (4) of $^{15}$C.

| $i$ | $\gamma_i$ | $C_i$ |
| --- | --- | --- |
| 1 | 4.400254682811078E-003 | –1.911899202003393E-003 |
| 2 | 1.080053848380744E-002 | –1.812966082174055E-002 |
| 3 | 2.564236030232376E-002 | –5.667440751622722E-002 |
| 4 | 5.871946600420144E-002 | –1.100036773570067E-001 |
| 5 | 1.272854480145382E-001 | –1.689491422977194E-001 |
| 6 | 2.722367794808272E-001 | –8.927563832717325E-002 |
| 7 | 3.451311294813448E-001 | 4.976170503854868E-001 |
| 8 | 5.153383537797701E-001 | 4.660987945089151E-001 |
| 9 | 7.170517147157914E-001 | 1.389709903982375E-001 |
| 10 | 1.048194029972669 | 5.167447028899066E-003 |

*Note.* The normalization factor of the wave function on the interval of 0–30 fm is $N$ = 9.999929970958931E-001.

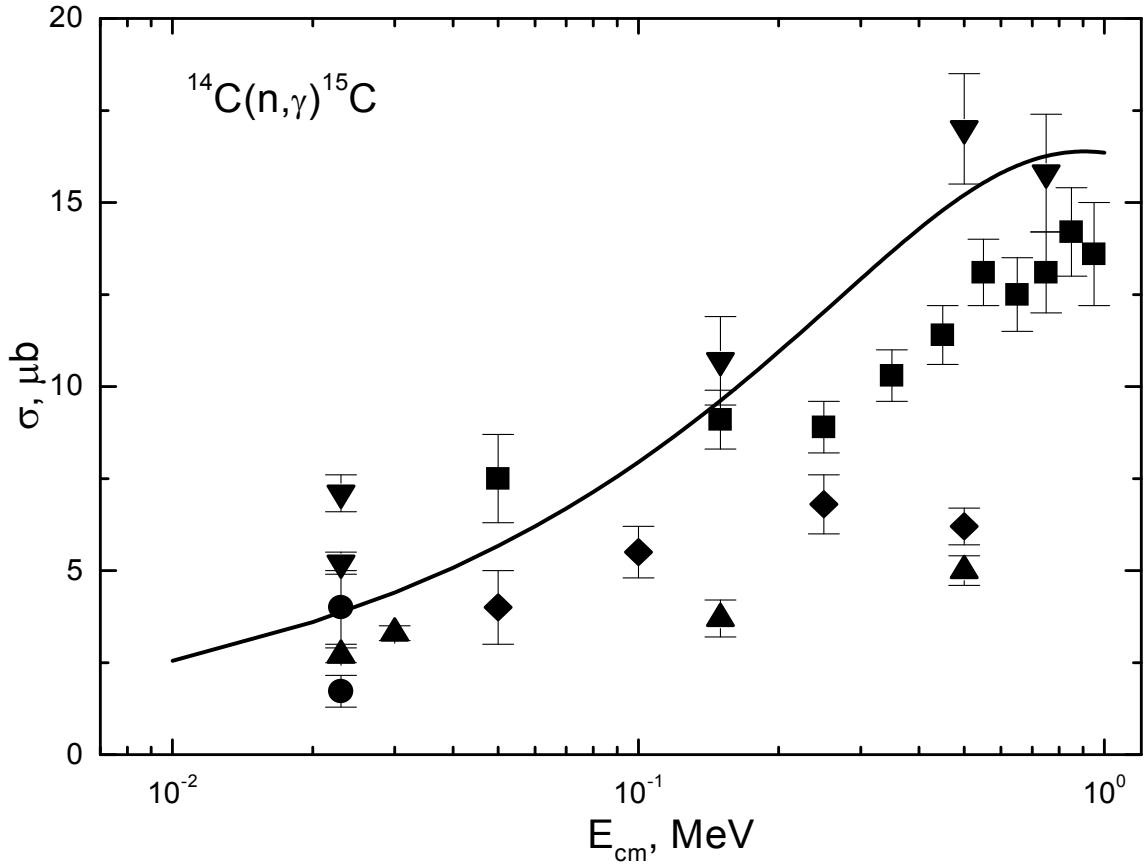

**Figure 1a:** The total cross sections of the radiative neutron capture on $^{14}$C. Squares (■) represent experiment data from [26], points (●) – from [27], triangles (▲) – from [28], reverse triangles (▼) – from [29], rhombs (♦) – from [30]. Lines show the total cross-sections calculation results.

Going over to the direct description of the results of our calculations, let us note that available experimental data for the total cross sections of the radiative neutron capture on $^{14}$C [26–30], found using the MSU data base [14], show the existence of big ambiguities of these cross sections, measured in different works. For example, the difference of cross sections at the energy 23 keV [27–29] is equal to two-three times, and ambiguity of the different data at the energy range 100–1000 keV reaches three-four times [26,28–30]. The experimental results at the energy range 23 keV–1.0 MeV for works mentioned above are shown in figures 1a and 1b. Here, we are considering only the $E$1 capture from the $^2P$ scattering states to the $^2S_{1/2}$ ground state of $^{15}$C, because the contribution of the $E$2 radiative capture process with the transition from the $^2P$ scattering waves to the first excited state with $J^\pi = 5/2^+$ that can be compare to the $^2D_{5/2}$ level is 25–30 times less as it was shown in work [31], and it is possible to neglect it in the presence of existent errors and ambiguities in measurements of total cross sections.

Thereby, we have considered only the $E$1 transition to the GS from the non-resonance $^2P_{1/2}$ and $^2P_{3/2}$ scattering waves at energies lower 1 MeV, with zero depth potential without FS, i.e., with zero scattering phase shifts. The total cross section calculation results of the radiative neutron capture on $^{14}$C with the GS potential of Eq. (4) given above, at energies lower 1 MeV, are shown in figures 1a and 1b by the solid lines, at that the calculation results are starting from the energy 1 eV. It is seen that these results are in a better agreement with data from [29]. Thereby, the total capture cross sections completely depend from the form of the potential of the ground state of $^{15}$C in the n$^{14}$C channel, because the $^2P$ potentials of the input channel without FS can be simply zeroized at the considered energies. The used GS potential, which allows us to acceptably describe the existent experimental data for the total radiative capture cross sections, leads to the correct description of the basic characteristics of the GS, specifically, the binding energy, the mean square radius and the AC of $^{15}$C in the n$^{14}$C channel.

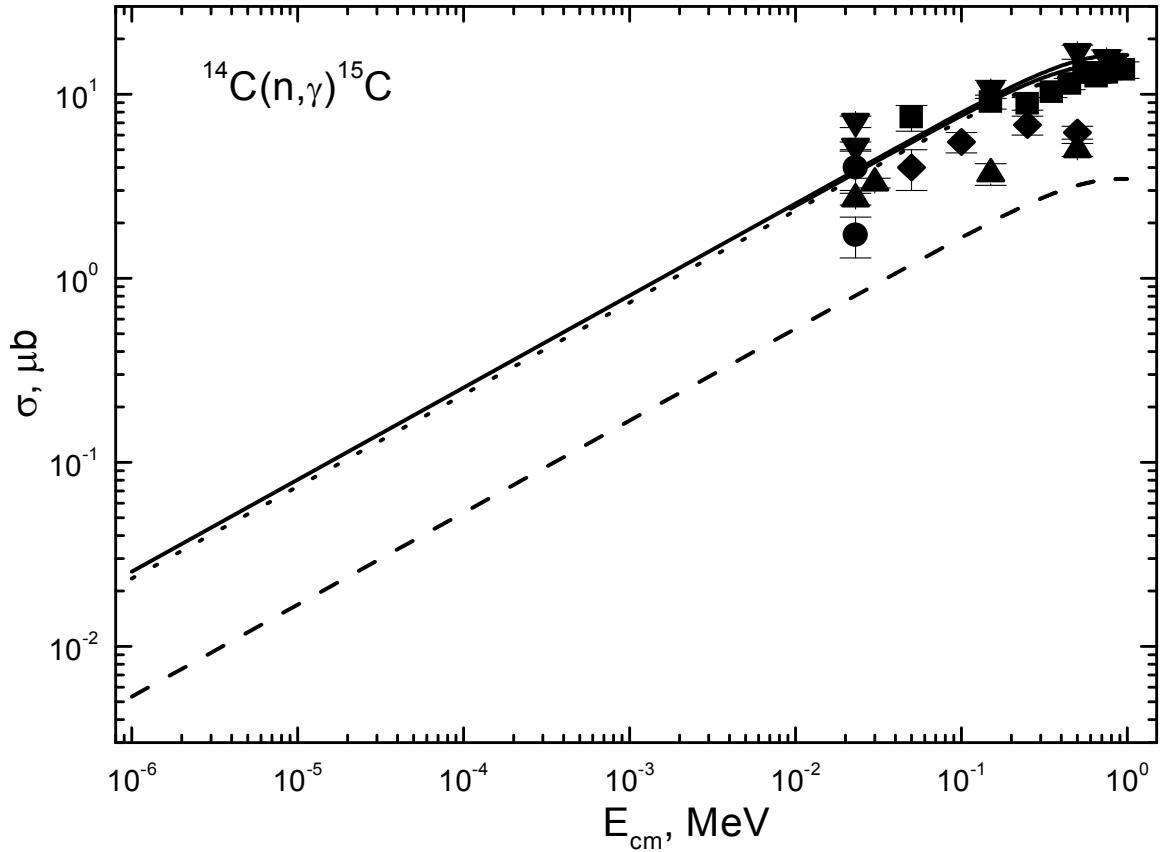

**Figure 1b:** The total cross sections of the radiative neutron capture on $^{14}$C. Squares (■) represent experiment data from [26], points (●) – from [27], triangles (▲) – from [28], reverse triangles (▼) – from [29], rhombs (♦) – from [30]. Lines show the total cross sections calculation results.

Besides, let us note that if we will use the GS potential of $^{15}$C without FS, for example, with parameters:

$V_{GS}$ = 19.994029 MeV and $\alpha_{GS}$ = 0.2 fm$^{-2}$,

leading to the binding energy of -1.218090 MeV, AC of 1.46(1) at the range 5–30 fm, charge and mass radiuses of 2.51 fm and 2.63 fm, correspondingly. The total cross section calculation results are shown in figure 1b by the dashed line, which is appreciably lower than all experimental data.

For the purpose of obtaining results that, in general, correctly describe experiment, we need parameters of the potential without FS:

$V_{GS} = 4.593639$ MeV and $\alpha_{GS} = 0.02$ fm$^{-2}$,

that leads to the big width and small depth of the interaction; the cross section calculation results are shown in figure 1b by the dotted line, which is practically superposed with the solid line. This potential leads to the charge radius of 2.53 fm, mass radius of 2.92 fm, and its AC at the interval 15–30 fm is equal to 3.24(1). We can see from these results that both variants of such GS potentials reproduce the AC value [25] incorrectly, although the last of them allows to reproduce the experimental total capture cross sections.

Because at energies from 1 eV up to 1 keV the calculated cross section is almost a straight line (see figure 1b, solid line), it can be approximated by a simple function of the form:

$$\sigma_{ap}(\mu b) = 0.7822\sqrt{E_n(\text{keV})}.$$

The value of given constant 0.7822 $\mu$b·keV$^{-1/2}$ was determined by a single point at a cross section with minimal energy of 1·eV. Furthermore, it turns out that the absolute value of the relative deviation of the calculated theoretical cross sections, and the approximation of this cross section by the expression given above at energies less than 1 keV, does not exceed 0.4%. Apparently, it can be suggested that this form of the total cross section energy dependence will be retained at lower energies. Therefore, we can perform an evaluation of the cross section value; for example, at the energy of 1 meV ($10^{-3}$ eV = $10^{-6}$ keV) – it gives result 0.78 $10^{-3}$ $\mu$b.

Thus, it is possible to describe the available experimental data only on the basis of the $E1$ transition from the $P$ scattering waves with zero potentials to the GS of $^{15}$C, which is described by the interactions that coherent with the main characteristics of this state – mean square radius, AC in the considered n$^{14}$C channel and two-particle binding energy. However, since for almost all earlier considered by us neutron capture processes on light nuclei, the total cross sections were measured down to 5–25 meV [15,19], then it is interesting to carry out the measurements of the cross sections for this reaction, even though for energy about 1 eV, when the given above formula forecast the value of 0.025 $\mu$b·

## 4. Total cross sections of the neutron capture on $^{14}$N

The ground state of $^{15}$N with $J^{\pi}, T = 1/2^-, 1/2$, since $^{14}$N has the moment $J^{\pi}, T = 1^+, 0$ [21], can be represented by the mixture of doublet $^2P_{1/2}$ and quartet $^4P_{1/2}$ states and, further, the $E1$ transitions at the non-resonance energy range to 0.5 – 0.6 MeV from the doublet $^2S_{1/2}$ and quartet $^4S_{3/2}$ scattering waves with one bound FS to the $^{2+4}P_{1/2}$ GS – $^2S_{1/2} + ^4S_{3/2} \rightarrow ^{2+4}P_{1/2}^{gs}$ will be considered as basic, i.e., the next total capture cross sections will be calculated:

$$\sigma_0(E1) = \sigma(E1, ^2S_{1/2} \rightarrow ^2P_{1/2}^1) + \sigma(E1, ^4S_{3/2} \rightarrow ^4P_{1/2}^1).$$

Thereby, as for the neutron capture on $^7$Li, the transitions to the doublet and quartet parts of wave functions of the ground state, which are not differ in this approach and correspond to the BS in the same potential. It should be noted here that there is excited level in the $^{15}$N spectrum at the energy 9.2221 MeV, but bound in the n$^{14}$N channel at -1.6112 MeV with $J^\pi = 1/2^-$ level. Thus, it is possible to analyze additional transitions $^2S_{1/2} + ^4S_{3/2} \to ^{2+4}P_{1/2}^{es}$, i.e., to consider total cross sections.

$$\sigma_1(E1) = \sigma(E1, ^2S_{1/2} \to ^2P_{1/2}^2) + \sigma(E1, ^4S_{3/2} \to ^4P_{1/2}^2) .$$

Here, we will limit ourselves to just transitions on the bound states with minimal values of $J^\pi = 1/2^\pm$. Therefore, let us consider further the possible transitions from the $P_{1/2}$ and $P_{3/2}$ scattering states to the second, seven and nine excited states (ES) of $^{15}$N with $J^\pi = 1/2^+$ at the energies 5.298822, 8.31262 and 9.04971 MeV, binding in the n$^{14}$N channel, which can be refer to the $^2S_{1/2}$ doublet wave.

Meanwhile, further we will consider the $^{2+4}P_{1/2}$ scattering state, which has the resonance at 492.6(0.65) keV and width about 8(3) keV (see, for example, Tables 15.4 and 15.14 of work [21] – the resonance at 11.2928(7) MeV). We will not take into account the resonance state placed near at 430(5) keV with $J^\pi \geq 3/2$ and width about 3 keV (energy 11.235(5) MeV in Table 15.4 [21]), we will not taking into account, for a while, certain parity [21]. The potentials of the $^{2+4}P_{3/2}$ waves will take equal to zero, because they have not forbidden states and resonances lower 1.0 MeV. So, then we can analyze processes $^2P_{1/2} + ^2P_{3/2} \to ^2S_{1/2}^{es}$ and represent total cross sections in the form:

$$\sigma_3(E1) = \sigma(E1, ^2P_{1/2} \to ^2S_{1/2}^1) + \sigma(E1, ^2P_{3/2} \to ^2S_{1/2}^1) +$$
$$\sigma(E1, ^2P_{1/2} \to ^2S_{1/2}^2) + \sigma(E1, ^2P_{3/2} \to ^2S_{1/2}^2) +$$
$$\sigma(E1, ^2P_{1/2} \to ^2S_{1/2}^3) + \sigma(E1, ^2P_{3/2} \to ^2S_{1/2}^3).$$

Proceeding to the construction of the potentials for all these states, let us note that doublet and quartet bound $S$ levels with one bound FS are evidently differ in binding energy and the interaction potentials will be obtained for each of them. The $^2S$ and $^4S$ scattering potentials with one bound FS are also evidently differ, but we did not succeed in construction of the potentials, which are able to take into account the resonances in these waves. Therefore, further we will consider that they have to result in the near-zero scattering phase shifts and we will use the same potential parameters for them. Let us note, that the first resonance in the $^2S_{1/2}$ wave is at the energy 11.4376(0.7) MeV with $J^\pi = 1/2^+$ or 0.639(5) MeV (c.m.) above threshold of the n$^{14}$N channel with the width of 34 keV in the laboratory system (l.s.), and in the $^4S_{3/2}$ wave at the energy 11.763(3) MeV with $J^\pi = 3/2^+$ or 0.998(5) MeV above threshold of the n$^{14}$N channel with the neutron width about 45 keV (l.s.) [21].

The $^{2+4}P$ scattering states or the $P$ bound levels are mixed by spin, because the total moment $J^{\pi} = 1/2^-$ or $J^{\pi} = 3/2^-$ can be obtained both for $^2P$ and for $^4P$ waves. Therefore, further potentials of the $^{2+4}P_J$ states are constructed for states of the total moment $J$ and also become mixed by spin.

Totally, the parameter values that do not take into account the presence of the resonances in the considered energy range were used for the potentials of the doublet $^2S_{1/2}$ and quartet $^4S_{3/2}$ scattering waves

$$V_S = -19.0 \text{ MeV}, \quad \gamma_S = 0.06 \text{ fm}^{-2}, \qquad (5)$$

and the calculation results of the $^2S_{1/2}$ and $^4S_{3/2}$ phase shifts with this potential at the energies up to 1.0 MeV lead to the values in the range $0\pm2°$.

The next parameters are used for the resonance potential at 493 keV of the $^{2+4}P_{1/2}$ wave without FS:

$$V_P = -13328.317 \text{ MeV}, \quad \gamma_P = 50.0 \text{ fm}^{-2}, \qquad (6)$$

which lead to the resonance energy 493 keV at the level width 18.2 keV, that slightly more than the measured value [21]. Let us note that such potential should be done narrower for obtaining a correct value of the level width, while the width parameter already has the unusually big value.

The potential of the $^{2+4}P_{1/2}$ state without FS has to correctly reproduce the binding energy of the GS of $^{15}$N in the n$^{14}$N channel at -10.8333 MeV [21] and reasonably describe mean square radius of $^{15}$N, which has the experimental value of 2.612(9) fm [21] at the experimental radius of $^{14}$N equals 2.560(11) [14]. Consequently, the next parameters for the GS potential of $^{15}$N in the n$^{14}$N channel without FS were obtained

$$V_{GS} = -55.442290 \text{ MeV}, \quad \gamma_{GS} = 0.1 \text{ fm}^{-2}. \qquad (7)$$

The potential leads to the binding energy of -10.83330001 MeV at the FDM accuracy equals $10^{-8}$ MeV, to the mean square charge radius $R_{ch} = 2.57$ fm and mass radius of 2.73 fm. The value of 4.94(1) was obtained for the asymptotic constant in the dimensionless form at the range 7–13 fm. The error of the constant is determined by its averaging over the above mentioned distance. The value 5.69(7) fm$^{-1/2}$ is given in works [32,33], and it gives 4.81(6) after recalculation to the dimensionless quantity at $\sqrt{2k} = 1.184$.

The variational method was used for the additional control of calculations of the GS energy, which with the dimension of the basis $N = 10$ and independent variation of the potential parameters for the potential of Eq. (7) allowed to obtain the energy of -10.83330000 MeV. The parameters of the variational radial WF are listed in Table 2 and residuals have the order of $10^{-8}$. The asymptotic constant at the range 7 – 14 fm is equal to 4.9(1), and the charge radius does not differ from the value obtained in above FDM calculations.

Consequently, the average value of -10.833300005(5) MeV can be taken as a realistic estimate of the binding energy in this potential. Therefore, the accuracy of determination of the binding energy of $^{15}$N for the given two-cluster potential of Eq. (7) in the two-particle channel, obtained by two methods (FDM and VM) and by two different computer programs, can be written as $\pm5\cdot10^{-9}$ MeV = $\pm5$ meV and it practically coincide with the given FDM accuracy.

**Table 2:** The coefficients and expansion parameters of the radial variational wave function of the n$^{14}$N system for the GS potential of Eq. (7) of $^{15}$N.

| i | $\gamma_i$ | $C_i$ |
|---|---|---|
| 1 | 2.763758363387135E-002 | -3.542736101468866E-004 |
| 2 | 5.252886535294879E-002 | -6.584466560462019E-003 |
| 3 | 9.444752983481111E-002 | -4.679747477384075E-002 |
| 4 | 1.088660404093062E-001 | 1.318491526218144E-002 |
| 5 | 1.503486884800729E-001 | -1.087314408835770E-001 |
| 6 | 2.226413018972464E-001 | -8.992256982354141E-002 |
| 7 | 2.684402252313877E-001 | -5.745985900990638E-002 |
| 8 | 3.736191607656845E-001 | -2.834463978909703E-002 |
| 9 | 7.499707281247036E-001 | -1.302287964205851E-004 |
| 10 | 6.009438691088970 | 2.060489845515652E-006 |

*Note.* The normalization factor of the wave function on the interval of 0–30 fm is N = 1.000000000000001.

The potential parameters of the ES of $^{15}$N at the energy -1.6112 in the n$^{14}$N channel with the moment $J^\pi$ = 1/2$^-$, coincided with the GS moment, have the values:

$$V_{ES} = -33.120490 \text{ MeV}, \quad \gamma_{ES} = 0.1 \text{ fm}^{-2}. \tag{8}$$

The potential leads to the binding energy of -1.611200 MeV at the FDM accuracy equals 10$^{-6}$ MeV, to the mean square charge radius $R_{ch}$ = 2.58 fm and the mass radius of 2.71 fm. The value of 1.19(1) was obtained for the asymptotic constant in the dimensionless form at the range 8–30 fm.

The same width as for the GS was used for potentials of ES with $J^\pi$ = 1/2$^+$ of $^{15}$N, which compare to the $^2S_{1/2}$ bound levels in the n$^{14}$N channel. For example, for the potential of the first bound in the n$^{14}$N channel $^2S_{1/2}$ state at the energy 5.298822 MeV relatively to the GS or -5.534478 MeV relatively to the threshold of the n$^{14}$N channel the next values are used:

$$V_{1S} = -66.669768 \text{ MeV}, \quad \gamma_{1S} = 0.1 \text{ fm}^{-2}. \tag{9}$$

The potential leads to the binding energy of -5.534478 MeV, to the charge radius of 2.57 fm, the mass radius of 2.69 fm and the value of 5.65(1) was obtained for the AC at the range 8–20 fm.

The next parameters were used for the potential of the second bound in the n$^{14}$N channel $^2S_{1/2}$ state at the energy of 8.31263 MeV relatively to the GS or -2.52068 MeV relatively to the threshold of the n$^{14}$N channel:

$$V_{2S} = -56.271191 \text{ MeV}, \quad \gamma_{2S} = 0.1 \text{ fm}^{-2}. \tag{10}$$

The potential leads to the binding energy of -2.520680 MeV, to the charge radius of 2.58 fm, the mass radius of 2.77 fm and the value of 3.34(1) was obtained for the

AC at the range 8–21 fm.

These parameters were used for the potential of the third bound in the n$^{14}$N channel $^2S_{1/2}$ state at the energy of 9.04971 MeV relatively to the GS or -1.78362 MeV relatively to the threshold of the n$^{14}$N channel:

$V_{3S}$ = -53.1402573 MeV,  $\gamma_{3S}$ = 0.1 fm$^{-2}$. (11)

The potential leads to the binding energy of -1.783620 MeV, to the charge radius of 2.58 fm, the mass radius of 2.82 fm and the value of 2.78(1) was obtained for the AC at the range 8–27 fm.

Proceeding to the direct description of the results of our calculations, let us note that all used experimental data for total cross sections of the radiative neutron capture on $^{14}$N were obtained from the data base [14,34], and the data itself are given in works [35-42]. These data for total cross sections of the radiative neutron capture on $^{14}$N have been obtained for the energy range 25 meV – 65 keV and are shown in figure 2 together with our calculation results (the solid line) of the total summarized cross sections of the radiative neutron capture on $^{14}$N to the considered above bound $^{2+4}P_{1/2}$ and $^2S_{1/2}$ states of $^{15}$N with the given potentials at energies lower than 1 MeV. The results of different works at energy 25 meV have the values of the cross sections at the range 77–80 mb. They are given in figure 2 by one point and, for example, in one of the latest work [41] the value of 80.3(6) mb was obtained for this energy.

The calculations of transition from the S scattering waves of Eq. (5) to the GS of Eq. (7) are shown by the dashed line in the non-resonance range, the dashed-dot line shows the transition to the excited $^{2+4}P_{1/2}$ state with the potential of Eq. (8), and the dashed-dot-dot line identifies their sum. The resonance part of cross sections for potential of the $^2P_{1/2}$ scattering in the form of Eq. (6) and zero potential of the $^2P_{3/2}$ wave of continuous spectrum is formed by the transitions to the first, second and third bound $^2S_{1/2}$ states with potentials of Eqs. (9)-(11). These cross sections are denoted by the dashed, dotted and dashed-dot lines, and their sum is shown by the solid line in the resonance range, i.e., near 5 – 1000 keV. The total summarized cross sections taking into account all considered transitions are also shown by the solid line at all energy range, i.e., from 10$^{-5}$ keV to 1.0 MeV.

It is seen from the given figure 2 that results of our calculations are quite well describe the data on capture cross sections at 25 meV, but do not reproduce measurements of work [42] at the energy 65 keV, given by open circle. The reason of this, apparently, is in the absence of taking into account transitions from the resonance $^2S_{1/2}$ and $^4S_{3/2}$ scattering waves with relatively big width to the GS and excited $^{2+4}P_{1/2}$ states with $J^\pi$ = 1/2$^-$.

Furthermore, if, for comparison, we will use the $^2S_{1/2}$ and $^4S_{3/2}$ scattering potentials with zero phases and zero depth, which have not FS, i.e., do not agree with the given above classification of FS and AS according to Young schemes, then the results of cross section calculation for potentials of the GS of Eq. (7) and ES of Eq. (8) lie above all of the experimental data more than by an order.

As it is seen from the obtained results, the description of the total radiative capture cross sections at lowest energies and BS characteristics, including AC of $^{15}$N in the n$^{14}$N channel can be well harmonized on the basis of the considered combination of the potentials of Eqs. (5) and (7). In other words, if to fix parameters of the GS potential of $^{15}$N in the n$^{14}$N channel on the basis of the correct description of its characteristics, including AC. Therefore, it is quite possible on the basis of classification of FS and AS according to Young schemes to find such $^{2+4}S_{1/2}$ scattering potentials that allow us correctly to describe not only near-zero elastic scattering phase shifts but also the value of cross sections of the radiative neutron capture on $^{14}$N at the energy of 25 meV. The future measurements of total cross sections for other energies would make it possible to clarify the quality of description of such cross sections in the considered model for potentials with FS more unambiguously.

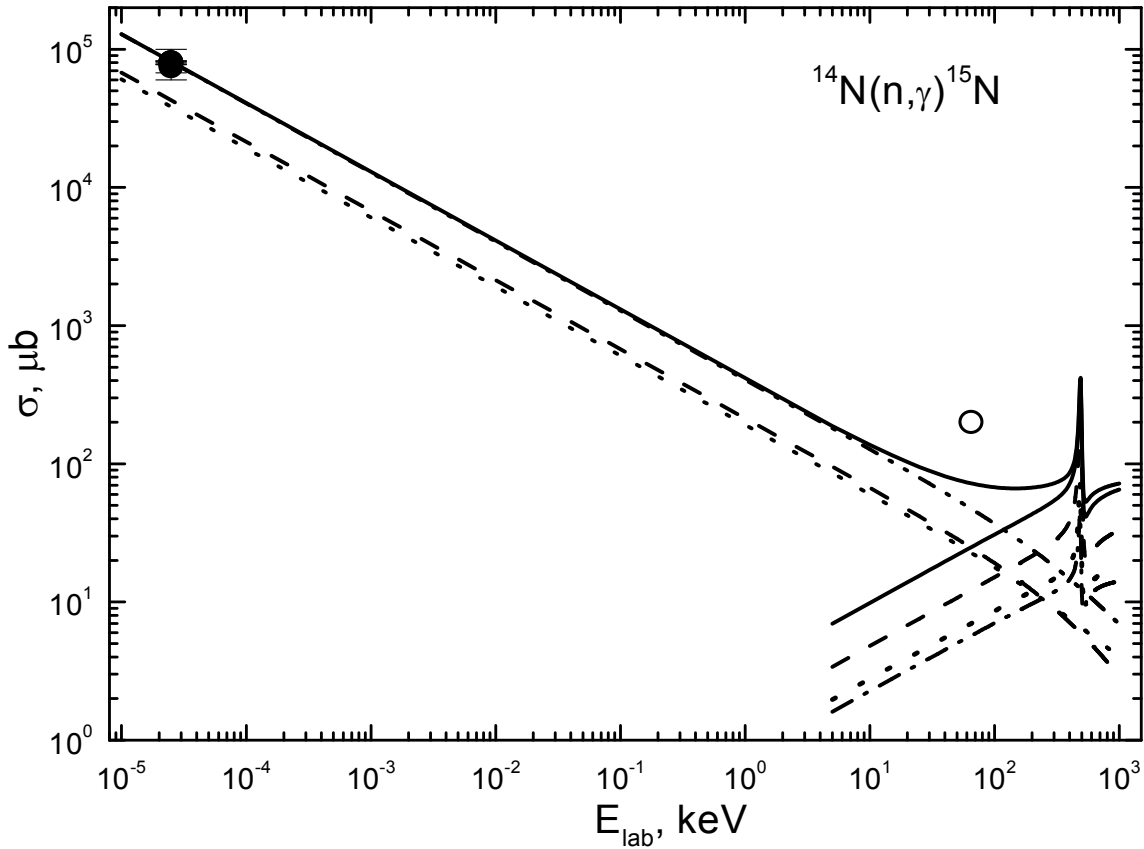

**Figure 2.** The total cross sections of the radiative neutron capture on $^{14}$N. Points (•) – represent experiment data from [35-41], open circles (○) – from [42]. Lines show the total cross sections calculation results given in the text.

Since at the energies from 10 meV to 10 keV the calculated cross section, given in figure 2 by the solid line. is almost straight line, it can be approximated by the simple function of energy from Eq. (2). The constant value 406.4817 μb·keV$^{1/2}$, in this case, was determined by a single point at cross-sections with minimal energy of 10 meV (l.s.). The absolute value from Eq. (3) of relative deviation of the calculated theoretical cross sections and the approximation of this cross section by the given above function from Eq. (2) in the range to 10 keV is less than 0.9%. If to perform an evaluation of the cross section value, for example, at the energy of 1 μeV ($10^{-6}$ eV = $10^{-9}$ keV), result is 12.8 b.

## 5. Conclusion

Thereby, it is possible to coordinate description of the elastic scattering processes

(phase shifts), main BS characteristics of nuclei (energy, radius, AC) and total radiative capture cross sections for all considered above processes of radiative capture on the basis of unified variants of intercluster potentials for each reaction. In addition, all these results are obtained on the basis of the general classification of orbital cluster states according to Young schemes [6,7,18,19].

Note, in conclusion, that there are already twenty cluster systems, considered by us earlier based on the potential cluster model with the classification of the orbital states according to the Young schemes, in terms of which it is possible to obtain acceptable results on description of the characteristics of the radiative neutron or light nuclei capture processes on atomic nuclei. Thereby, we have used the intercluster potentials agreed with the scattering phase shifts of the considering clusters in the input channel and with the BS characteristics of nuclei that result from reaction. The properties of these cluster nuclei and some characteristics of their BS in the considering cluster channel are given in Table 3.

**Table 3:** The characteristics of nuclei and cluster systems (spin, isospin and parity), and references to works where the astrophysical *S*-factors and total cross sections of the radiative capture reactions in which they were considered on the basis of potential cluster model.

| No. | Nucleus ($J^{\pi}, T$) | Cluster channel | $T_z$ | $T$ | Ref. |
|---|---|---|---|---|---|
| 1. | $^3$H ($1/2^+$, $1/2$) | n$^2$H | $-1/2 + 0 = -1/2$ | $1/2$ | [19] |
| 2. | $^3$He ($1/2^+$, $1/2$) | p$^2$H | $+1/2 + 0 = +1/2$ | $1/2$ | [6, 7] |
| 3. | $^4$He ($0^+$, $0$) | p$^3$H | $+1/2 - 1/2 = 0$ | $0 + 1$ | [6, 7] |
| 4. | $^6$Li ($1^+$, $0$) | $^2$H$^4$He | $0 + 0 = 0$ | $0$ | [8] |
| 5. | $^7$Li ($3/2^-$, $1/2$) | $^3$H$^4$He | $-1/2 + 0 = -1/2$ | $1/2$ | [8] |
| 6. | $^7$Be ($3/2^-$, $1/2$) | $^3$He$^4$He | $+1/2 + 0 = +1/2$ | $1/2$ | [8] |
| 7. | $^7$Be ($3/2^-$, $1/2$) | p$^6$Li | $+1/2 + 0 = +1/2$ | $1/2$ | [6, 7] |
| 8. | $^7$Li ($3/2^-$, $1/2$) | n$^6$Li | $-1/2 + 0 = -1/2$ | $1/2$ | [19] |
| 9. | $^8$Be ($0^+$, $0$) | p$^7$Li | $+1/2 - 1/2 = 0$ | $0 + 1$ | [6, 7] |
| 10. | $^8$Li ($2^+$, $1$) | n$^7$Li | $-1/2 - 1/2 = -1$ | $1$ | [19] |
| 11. | $^{10}$B ($3^+$, $0$) | p$^9$Be | $+1/2 - 1/2 = 0$ | $0 + 1$ | [6, 7] |
| 12. | $^{10}$Be ($0^+$, $1$) | n$^9$Be | $-1/2 - 1/2 = -1$ | $1$ | In print |
| 13. | $^{13}$N ($1/2^-$, $1/2$) | p$^{12}$C | $+1/2 + 0 = +1/2$ | $1/2$ | [6, 7] |
| 14. | $^{13}$C ($1/2^-$, $1/2$) | n$^{12}$C | $-1/2 + 0 = -1/2$ | $1/2$ | [15] |
| 15. | $^{14}$N ($1^+$, $0$) | p$^{13}$C | $+1/2 - 1/2 = 0$ | $0 + 1$ | [16] |
| 16. | $^{14}$C ($0^+$, $1$) | n$^{13}$C | $-1/2 - 1/2 = -1$ | $1$ | [15] |
| 17. | $^{15}$C ($1/2^+$, $3/2$) | n$^{14}$C | $-1/2 - 1 = -3/2$ | $3/2$ | This paper |
| 18. | $^{15}$N ($1/2^-$, $1/2$) | n$^{14}$N | $-1/2 + 0 = -1/2$ | $1/2$ | This paper |
| 19. | $^{16}$N ($2^-$, $1$) | n$^{15}$N | $-1/2 - 1/2 = -1$ | $1$ | [19] |
| 20. | $^{16}$O ($0^+$, $0$) | $^4$He$^{12}$C | $0 + 0 = 0$ | $0$ | [6, 7] |

**Acknowledgments**

This work was supported by the Grant Program No. 0151/GF2 of the Ministry of Education and Science of the Republic of Kazakhstan: The study of thermonuclear processes in the primordial nucleosynthesis of the Universe.

We would like to express our thanks to Professor R. Yarmukhamedov for the detailed discussions of some questions of the work and for the provision of his results on the asymptotic normalization constants.